\documentclass{article}
\usepackage[utf8]{inputenc}
\usepackage{amsmath,amssymb}
\usepackage{amsfonts}
\usepackage{enumitem}
\usepackage[margin=1in]{geometry}
\usepackage{natbib}
\usepackage{hyperref}
\usepackage{graphicx}
\usepackage{soul}
\usepackage{setspace}
\usepackage{listings}
\usepackage{color,soul}
\usepackage{bbm}
\usepackage{authblk}
\usepackage[title]{appendix}
\usepackage{pdfpages}
\newcommand{\indep}{\perp \!\!\! \perp}
\newcommand{\sgn}{\operatorname{sgn}}
\newtheorem{definition}{Definition}

\newlist{condenum}{enumerate}{1} % 'condenum': a new, enumerate-like list env.
\setlist[condenum]{label=\bfseries Condition \arabic*., ref=\arabic*, wide}

\title{Methods for differential network estimation:\\ an empirical comparison}

\author[1*]{Anna Plaksienko}
\author[1]{Magne Thoresen}
\author[2]{Vera Djordjilović}
\affil[1]{Oslo Centre for Biostatistics and
Epidemiology, Department of Biostatistics,
University of Oslo, Oslo, Norway}
\affil[2]{Department of Economics, Ca’ Foscari University of Venice, Venice, Italy}
\affil[*]{\textit{email: annapla@uio.no}}

\begin{document}
\date{December 2024}

\maketitle

\begin{abstract}
We provide a review and a comparison of methods for differential network estimation in Gaussian graphical models with focus on structure learning. We consider the case of two datasets from distributions associated with two graphical models. In our simulations, we use five different methods to estimate differential networks. We vary graph structure and sparsity to explore their influence on performance in terms of power and false discovery rate. We demonstrate empirically that presence of hubs proves to be a challenge for all the methods, as well as increased density. We suggest local and global properties that are associated with this challenge. Direct estimation with lasso penalized D-trace loss is shown to perform the best across all combinations of network structure and sparsity.
\end{abstract}


\section{Introduction}

Nowadays, networks are used in many different areas of science in various forms, capturing different types of interactions between variables. With the rise of situations when datasets come from similar yet distinct conditions, it becomes more relevant to estimate the differences between those conditions, specifically in terms of networks. One example can be a molecular network of a healthy population compared to cancer patients: many genes are unrelated to the disease, so their connections would be identical in both groups. It might be more relevant to estimate the differential network, i.e., only the connections that are different between conditions. In this article, we highlight possible challenges in such differential network estimation. To the best of our knowledge, such issues were not previously discussed in the literature specifically for differential graph estimation. Therefore, we hope this manuscript may be helpful to others pursuing this field. 

We refer the reader to \citep{Shojaie_review} for the broad review of differential network estimation. Here, we define a narrower setting and perform a comparison of methods there. 

Broadly, a network is a collection of vertices corresponding to some variables and edges, signifying some interaction between these variables that may be defined in various ways. In our work, we focus on graphical models – networks where edges signify conditional dependence, i.e., direct, unmediated connections between variables (see the next section for the formal definition and \citep{Lauritzen} for in-depth details). Although networks defined through marginal associations are relevant and of interest in many applications, we do not discuss them here.

Within graphical models, we focus specifically on undirected ones. Firstly, directed graphical models are more challenging to estimate \citep{DAGs} and additional constraints are required to ensure identifiability \citep{Giraud2015IntroductionStatistics}. Secondly, in many cases, it is of interest to focus only on the presence of a connection, not necessarily its direction.

In the class of undirected graphical models, we consider those associated with the Gaussian distribution. As in many other situations, this particular distribution has many unique theoretical properties that make its use beneficial. In the case of graphical models, it is the fact that the Gaussian graph structure can be estimated through a precision (inverse covariance) matrix. 

In addition, we assume that data comes in the form of two datasets from similar but different conditions. It can be measurements of the same variables processed on different equipment or in different locations, the same system under various stimulating conditions, different subtypes of diseases, possibly different time points. We assume that there are edges (connections) that coincide in the graphical models  from two  conditions, but that there are also differences  and we aim to capture those.

In this manuscript, we apply a frequentist approach and do not discuss Bayesian methods, although many exist (see, for example \citep{Bayesian_evaluation}). This choice was made mainly for better comparability of the methods. 

We include methods both with and without error control in the comparison, in order to more comprehensively assess various approaches. However, we would like to emphasize the importance of error control, especially in biological settings.

Most standard methods for estimation of Gaussian graphical models (like the graphical lasso) implicitly assume a uniformly random network, and hence, so do methods for differential network estimation. However, there is clear evidence that this might not be a realistic enough model for biological networks. Thus, it becomes important to investigate to what extent these methods are able to handle other structures as well.

The aim of the article is twofold. First, we will perform an evaluation and a comparison of available methods for differential network estimation through a simulation study. Our focus is on structure learning, so we measure performance by power and false discovery rate (FDR) related to edge detection. Next, we will investigate the influence of graph structure on methods performance through the same simulation study.

\bigskip
We will start with the various definitions of a differential network. Next, we will describe the estimation methods and compare them in the simulation study, highlighting their advantages and downsides. We will discuss how different graph structures influence the performance of the methods as a class and each of them individually.
Furthermore, we will demonstrate the performance of the six methods in an applied data example, highlighting the variability of result, and in the end, we will discuss our findings and point to possible directions for future work.

\bigskip
Throughout the paper, matrices are denoted by bold capital letters $\mathbf{M}$. Their scalar entries are denoted with non-bold letters with two lower case indices, e.g. $M_{ij}$. Bold lower case letters denote vectors, i.e. $\boldsymbol{\mu}$. Bold italic upper case letters denote random vectors, e.g. $\boldsymbol{X}$, and non-bold italic upper case letters with one lower index denote random variables, e.g. $V_i$. Non-bold lower case letters with or without a lower index denote scalars, i.e. $\lambda_2$. 

\subsection{What is a differential network?} \label{sec: diff net}

First, we would like to highlight that multiple definitions of differential networks exist. Let us first define a Gaussian graphical model. We start with a $p$-variate zero mean Gaussian distribution of variables $\boldsymbol{X}=(X_1, \dots, X_p)^T$ with a covariance matrix $\mathbf{\Sigma}$. Note that we assume that the mean vector is zero without loss of generality, since the covariance structure, modeled by the graphical model, is structurally independent from the mean vector.  We can represent this distribution $\mathcal{N}(0, \mathbf{\Sigma})$ in terms of the graph $G = (V, E)$, where $V = \{1, \dots, p\}$ is the set of vertices and $E \subseteq V \times V$ is the set of edges. Each vertex $i$ corresponds to a random variable $X_i$, and the absence of an edge between vertices $i$ and $j$ implies conditional independence of the corresponding variables, i.e., $(i, j) \notin E$ implies that $ X_i \indep X_j | \boldsymbol{X}_{ V \setminus\left\{i,j\right\}}$, where ${\boldsymbol X}_U$ is a subvector of $\boldsymbol{X}$ for any $U\subset V$. In the Gaussian setting, the support of the precision matrix $\mathbf{\Omega} = \mathbf{\Sigma}^{-1}$ encodes the graphical structure. Let $\mathbf{A}$ denote the adjacency matrix of graph G, then $A_{ij} = 0$ if and only if  $\Omega_{ij} = 0$.

Now let us consider two distributions instead of one, $\mathcal{N}(0, \mathbf{\Sigma}^{(1)})$ and $\mathcal{N}(0, \mathbf{\Sigma}^{(2)})$. They can describe two conditions, such as subtypes of a disease or data collected on different equipment or in different laboratories. In this case, it is of interest to investigate the differences between the two distributions. We can do so in terms of graphs again, as supports of precision matrices $\mathbf{\Omega}^{(1)}$ and $\mathbf{\Omega}^{(2)}$ encode graphs' $G^{(1)}$ and $G^{(2)}$ and  adjacency matrices $\mathbf{A}^{(1)}$ and $\mathbf{A}^{(2)}$. However, how do we define a differential network?

\begin{figure*}[h!]
\centering
  \includegraphics[width = 0.8\textwidth]{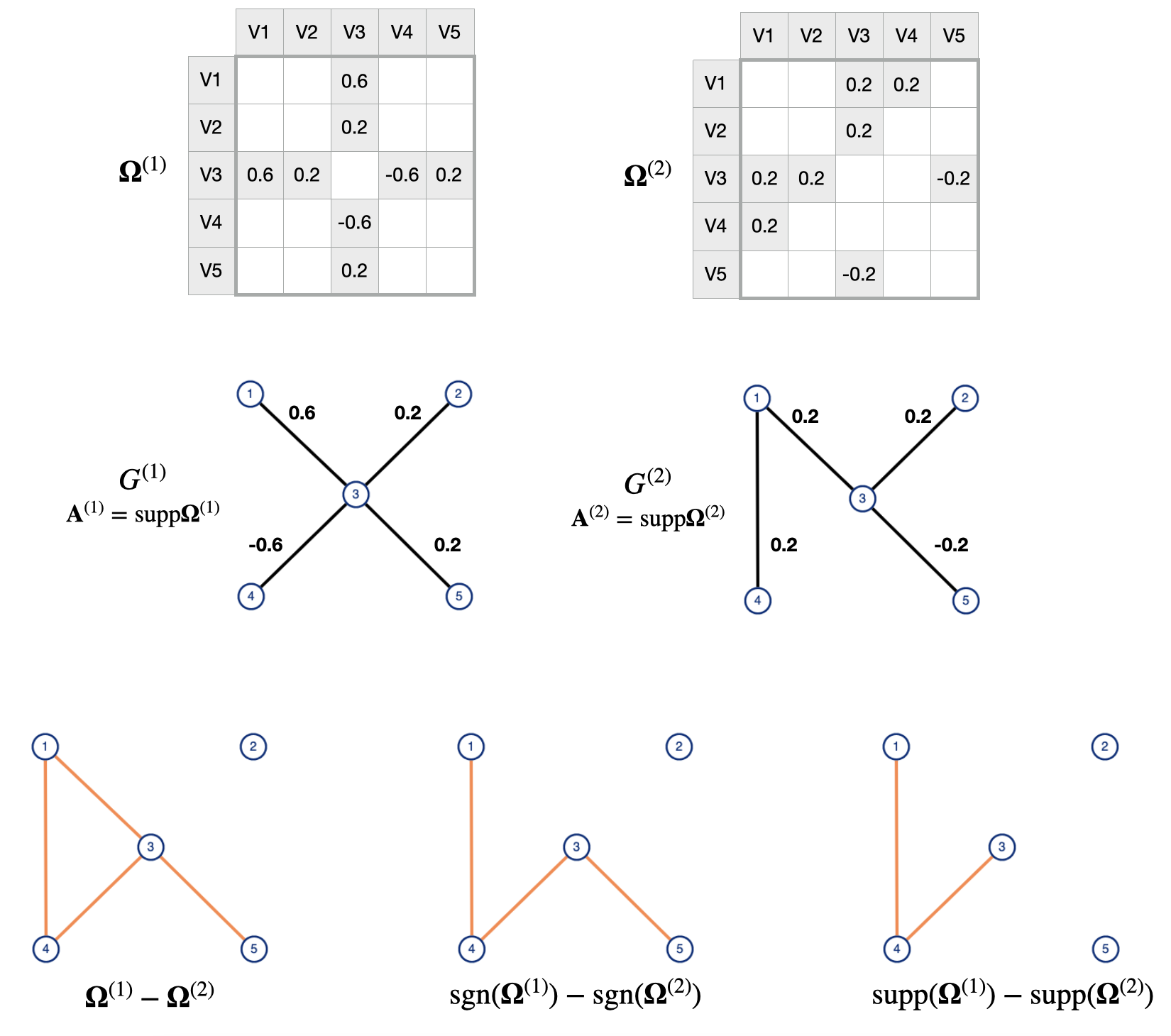}
  \caption{Illustration of different  definitions of differential networks. Top: precision matrices for two populations (blank squares represent zero entries). Center:  graphs of the two Gaussian graphical models. Bottom: differential networks based on differences in values (left), differences in sign (center) and differences in support (right). }
\label{fig_diff_definition}
\end{figure*}

The difference between two networks can be assessed in a number of ways. We report the three most prevalent definitions of a differential network within the Gaussian graphical models context. Let $G^{diff} = (V, E^{diff})$ denote a differential graph. Its edge set $E^{diff}$  can be defined as follows.

\begin{definition}[Difference in value]
\begin{equation}
E^{diff} = \left\{(i,j): \Omega^{(1)}_{ij} \neq \Omega^{(2)}_{ij}\right\}.
\end{equation}
\end{definition}

Definition 1 identifies as different, node pairs for which the corresponding elements of the two precision matrices are different. It captures quantitative differences between the two networks and is the most prevalent definition used in the literature. A difference in value may reflect either  a difference in structure or a difference in the strength of a relationship between the associated variables.  As an alternative, Definition 2 is  capturing qualitative differences. It identifies as different, node pairs for which the sign of the corresponding elements of the two precision matrices is different.

\bigskip

\begin{definition}[Difference in sign]
\begin{equation}
E^{diff} = \left\{(i,j): \sgn \left(\Omega^{(1)}_{ij}\right) \neq \sgn\left(\Omega^{(2)}_{ij}\right)\right\},
\end{equation}
where $\sgn(x) = 1$ for $x > 0$, $\sgn(x) = -1$ for $x < 0$, and $\sgn(x) = 0$ otherwise. 
\end{definition}

Definition 2 is thus capturing differences in structure and a subset of differences in values: those corresponding to a sign switch. Finally, Definition 3 captures solely differences in  structure.

\bigskip

\begin{definition}[Difference in support]
\begin{equation}
E^{diff} = \left\{(i,j): A^{(1)}_{ij} \neq A^{(2)}_{ij}\right\}.
\end{equation}
\end{definition}

Figure \ref{fig_diff_definition} illustrates the three definitions of a differential network. The choice of which definition to use in practical applications should be guided by subject matter considerations. Most methods reviewed here use Definition 1, with the exception of differential connectivity analysis (DCA), that uses Definition 3. Our simulation studies are designed so that differential networks according to the three definitions coincide. In this way, each method is used to estimate the type of difference it was originally intended for.

\section{Methods}

In this section, we present methods for estimating a differential network. We discuss only methods that provide an estimate of a differential network, that is, we do not consider methods that perform global testing of equality of two Gaussian distributions without estimating the differential structure itself.

Methods for estimating a differential network can be divided into three categories, as follows.

 \paragraph{\bf Joint estimation of multiple Gaussian graphical models.} These methods were designed to estimate multiple related Gaussian graphical models that are expected to have a partially common structure while exhibiting some differences. The output consists of the two condition specific networks, and it is up to the user to,  given the results, construct a differential network. We consider one method in this category: joint graphical lasso with fused penalty or, for short, the fused graphical lasso (FGL) \citep{JGL}.
\paragraph{\bf Testing-based methods.} These methods obtain an estimate of a differential network by testing equality of the entries of a partial correlation matrix \citep{TestingPartCorr}, a precision matrix \citep{TestingPrecMatr, ONDSA}, or an adjacency matrix \citep{DCA} under two conditions.   
\paragraph{\bf Direct estimation.} These methods estimate a differential network directly, without estimating condition specific networks \citep{Zhao2014, Dtrace}. We will discuss one such method, with lasso penalized D-trace loss \citep{Dtrace}.

\bigskip
In our simulation study, we consider representatives of each category. Our choice of specific methods was mainly guided by the software availability: we included only methods that have an available implementation in R.
An overview of the considered methods is given in Table \ref{table_overview} with a brief description of each.

\begin{table*}[h]
\footnotesize
\caption{\label{table_overview} Summary of methods for estimation of differential networks.}
\begin{tabular}{lllllll}
Name & Ref. & Type & Estimand & Error control & R package & Tun. param. \\
\hline
FGL & \citep{JGL} & Joint estim. & $\mathbf{\Omega}^{(1)}, \mathbf{\Omega}^{(2)}$ & No & \href{https://cran.r-project.org/package=JGL}{JGL} & $\lambda_1, \lambda_2$\\
PCor &  \citep{TestingPartCorr} & Testing&  $E
^{diff} = \left\{(i,j): \rho_{ij\cdot}^{(1)}\neq \rho_{ij\cdot}^{(2)}\right\}$ & Asymp. FDR  & \href{https://github.com/Zhangxf-ccnu/DiffNetFDR}{DiffNetFDR}& No\\
PMat &  \citep{TestingPrecMatr} & Testing  & $E
^{diff} = \left\{(i,j): 
\Omega_{ij}^{(1)}\neq \Omega_{ij}^{(2)}\right\}$ & Asymp. FDR & \href{https://github.com/Zhangxf-ccnu/DiffNetFDR}{DiffNetFDR}& No\\
ONDSA &  \citep{ONDSA} & Testing & $E
^{diff} = \left\{(i,j): 
\Omega_{ij}^{(1)}\neq \Omega_{ij}^{(2)}\right\}$& Asymp. FDR & \href{https://github.com/jiachenchen322/ONDSA}{ONDSA} & No\\
DCA & \citep{DCA} & Testing & $E^{diff} = \left\{(i,j): A_{ij}^{(1)} \neq A_{ij}^{(2)}\right\}$& Asymp. FDR & \href{https://github.com/sen-zhao/DCA}{DCA}& No\\
DTrace & \citep{Dtrace} & Direct estim. &  $\mathbf{\Omega}^{(1)} - \mathbf{\Omega}^{(2)}$ & No &\href{https://github.com/Zhangxf-ccnu/DiffGraph}{DiffGraph} & $\lambda$ \\
\hline
\end{tabular}
\end{table*}

We use the following notation. Suppose we are given two datasets $\mathbf{X}^{(k)}$, $k = 1, 2$, each with $n_k$ independent, identically distributed observations from $\mathcal{N}(0, \mathbf{\Sigma}^{(k)})$. We denote empirical covariance matrices as $\mathbf{S}^{(k)} = \dfrac{1}{n_k} \,\left({\mathbf{X}^{(k)}}\right)^{\top}\mathbf{X}^{(k)}$. We note that some of the presented methods consider $K > 2$ datasets but for simplicity and comparability we will consider only the case of $K = 2$. 

\subsection{Fused graphical lasso (FGL)} 

The fused graphical lasso \citep{JGL} is a likelihood-based method for the joint estimation of multiple graphical models from datasets with observations belonging to distinct classes. Although it is not the first method in this class – an earlier example can be \citep{Guo} – it is currently one of the most mentioned in the literature. The aim is to estimate two graphical models under the assumption that both networks are sparse, while at the same time enforcing some similarity between the networks. In the high-dimensional case of $p$ bigger than $n_k$, $k = 1, 2$, the authors suggest to maximize the following penalized log-likelihood function:
\begin{equation}\label{eq_FGL}
    \begin{split}
    \sum_{k=1}^2 &n_k \left\{ \log(\det \mathbf{\Omega}^{(k)}) - \mbox{tr}\left(\mathbf{S}^{(k)} \mathbf{\Omega}^{(k)}\right) \right\} -\\ &\lambda_1 \sum_{k=1}^2 \sum_{i \neq j} |\Omega_{ij}^{(k)}| - \lambda_2  \sum_{i, j} |\Omega_{ij}^{(1)} - \Omega_{ij}^{(2)}|,
    \end{split}
\end{equation}
where $\lambda_1$ and $\lambda_2$ are non-negative tuning parameters. The penalty term corresponding to $\lambda_1$ drives the sparsity of the estimated precision matrices, while the term corresponding to $\lambda_2$ encourages equality of entries across classes. The solution of  \eqref{eq_FGL} is called the fused graphical lasso (FGL).

In the paper, the authors suggest to construct a differential network according to the difference in  the entries of the estimated precision matrices, i.e. when $\hat{\Omega}^{(1)}_{ij} \neq \hat{\Omega}^{(2)}_{ij}$ (Definition 1). However, we will follow the approach from their simulation code and account for possible computational errors, defining an  edge $(i,j)$  as differential if $|\hat{\Omega}_{ij}^{(1)} - \hat{\Omega}_{ij}^{(2)}| > 10^{-3}$.

We use the authors' R package \texttt{JGL} available on CRAN. 

\subsection{Testing the equality of partial correlations (PCor)}

The authors of \citep{TestingPartCorr} defines a differential network through partial correlations. A partial correlation coefficient of $X_i$ and $X_j$ given $X_{l \ne i, j}$ can be expressed through precision matrix entries as $
    \rho_{ij\cdot} = -\Omega_{ij}/\sqrt{\Omega_{ii}\Omega_{jj}}.$
For two conditions, their differential graph is defined similarly to Definition 1 as
\begin{equation*}
    E^{diff} = \{(i, j) : \rho_{ij\cdot}^{(1)} \neq  \rho_{ij\cdot}^{(2)} \}.
\end{equation*}

The differential graph is estimated through a multiple testing procedure testing a collection of null hypotheses $\left\{H_{ij}:  \rho_{ij\cdot}^{(1)} = \rho_{ij\cdot}^{(2)}, 1 \le i < j \le p\right\}$. The authors prove that the proposed procedure provides asymptotic false discovery rate control, and further  recover an approximation of the set of common edges, again with false discovery rate control. We refer interested readers to the original article for further details.

In our simulation study, we used the implementation from the \texttt{DiffNetFDR} R package available on Github, which was not developed by the authors of the method, see software article \citep{DiffNetFDR_package} for further details. 

\subsection{Testing the equality of precision matrices (PMat)}

In \citep{TestingPrecMatr}, a differential network is defined according to Definition 1, as the difference between entries of precision matrices. The authors  first perform a  test of a global  null hypothesis $H_0: \mathbf{\Omega}^{(1)} = \mathbf{\Omega}^{(2)}$ and if it is rejected, investigate the structure of the differential network with a multiple testing procedure for a collection of null hypotheses $\left\{H_{ij}: \Omega_{ij}^{(1)} = \Omega_{ij}^{(2)}, 1 \le i < j \le p\right\}$. To perform the test of $H_{ij}$, the authors estimate  $\mathbf{\Omega}^{(k)}$  by  its relation with the coefficients of a set of node-wise linear regression models for $\mathbf{X}^{(k)}$. The test statistics are then obtained based on a bias-corrected estimator of the covariances between the residuals of the fitted models.

There is a Matlab implementation by the method's authors. In our simulation study, we used the implementation from the \texttt{DiffNetFDR} R package available on Github \citep{DiffNetFDR_package}. 

For other methods testing the equality of precision matrices entries, we refer the reader to \citep{He2019}, \citep{Belilovsky2016BrainConnectivity} (estimates individual precision matrices with debiased lasso) or \citep{SDA} (uses a symmetrized data aggregation strategy). 

\subsection{Omics Networks Differential and Similarity Analysis (ONDSA)}

In \citep{ONDSA}, a differential network is also defined by the differences in precision matrices (Definition 1) and differential and similar structures are identified in a two-step procedure with FDR control. First, individual networks are estimated via \citep{Ren2015}. In order to estimate individual entries $\mathbf{\Omega}_{ij}$,  ($X_i$, $X_j$) is regressed on the remaining variables by the scaled lasso. The scaled lasso is "tuning-free", as the authors suggest to fix $\lambda = \sqrt{(2/N) \log(p / \sqrt{N})}$. The residuals from this regression is then used to estimate $\mathbf{\Omega}_{ij}$. Then, the null hypothesis $H_0: \mathbf{\Omega}^{(1)}_{ij} = \mathbf{\Omega}^{(2)}_{ij}$ is tested for each edge to construct the set of differential edges $E^{diff}$. To correct for correlations among entry-wise test statistics caused by overlapping nodes in the network, a conditional FDR adjustment is applied, following the recommendation of \citep{Efron_2}. Furthermore, on the set complementary to the estimated differential edges $({E^{diff}})^c$, this method applies another testing procedure to divide the remaining edges into common nonzero edges and common zero edges (absence of edges).

In our simulations, we used the authors' package \texttt{ONDSA}, available on Github.

\subsection{Differential connectivity analysis (DCA)}

The authors of \citep{DCA} define a differential network through differences in structure of individual networks, i.e., differences in supports of precision matrices (Definition 3). Therefore, they suggest testing  qualitative differences,  rather than quantitative differences like the methods described above. The method is based on testing a collection of null hypotheses 
\begin{equation*}
    \left\{H_{i}: \mbox{ne}^{(1)}_i = \mbox{ne}^{(2)}_i, 1\leq i\leq p\right\},
\end{equation*}    
where $\mbox{ne}_i^{(k)}$ is the set of neighbours of node $i$ in $G^{(k)}$, i.e. $\mbox{ne}_i^{(k)}=\left\{j: (i,j) \in E^{(k)}\right\}$. To test this collection, a two-step procedure is proposed. First, two condition specific networks are estimated separately by any procedure that satisfies certain coverage conditions. In their manuscript, the authors use regression-based neighbourhood estimation \citep{MB2006}. For each node,  an estimate of a set of common neighbors  $\widehat{\mbox{ne}}_i^{0}$ is obtained as an intersection of the two condition specific neighborhoods: $\widehat{\mbox{ne}}_i^{0}=\widehat{\mbox{ne}}_i^{(1)} \cap \widehat{\mbox{ne}}_i^{(2)}$. In the second step, it is tested that there is no node $j$,  such that $j \notin \widehat{\mbox{ne}}_i^{0}$ and $(i,j) \in \widehat{\mbox{ne}}_i^{(1)} \cup \widehat{\mbox{ne}}_i^{(2)}$, i.e. there is no $j$ such that the edge $(i,j)$ is not a common edge, but it is present in one of the two estimated condition specific networks. 

There is a challenge of ''double-dipping'' – the same data is used both to formulate and to test the hypotheses. To address this issue, the authors adopt a data splitting approach. In their simulation study, they compare it with the naïve approach, which  treats hypotheses as data-independent. Not surprisingly, the naive approach was superior in statistical power, but, somewhat surprisingly, also in controlling false discovery rate. This was explained by the fact that the crucial assumption for controlling false discovery rate asymptotically is that the true common neighbourhood of each node is  covered by its estimator with probability tending to 1. This event is less likely to happen with smaller sample sizes that arise with sample splitting.

In our simulation study, we used the authors' R package \texttt{DCA} available on Github (not to be confused with the package of the same name on CRAN). We have added correction for multiplicity with the Benjamini-Hochberg procedure.

\subsection{Direct estimation with lasso penalized D-trace loss (DTrace)}

Here, we introduce an approach originally proposed in \citep{Zhao2014} and improved in \citep{Dtrace}. If $\mathbf{S}^{(k)}$, $k = 1, 2$, denotes a sample covariance matrix, it is suggested to estimate $\mathbf{\Delta} = \mathbf{\Omega}^{(1)} - \mathbf{\Omega}^{(2)}$ directly by minimizing a convex loss function
\begin{equation*}
    L_D(\mathbf{\Delta}, \mathbf{S}^{(1)}, \mathbf{S}^{(2)}) + \lambda|{\mathbf{\Delta}}||_1,
\end{equation*}
where
\begin{equation*}
\begin{split}
    L_D({\mathbf{\Delta}}, \mathbf{S}^{(1)}, \mathbf{S}^{(2)}) &= \dfrac{1}{4} \left(\langle\mathbf{S}^{(1)} \mathbf{\Delta}, \mathbf{\Delta}\mathbf{S}^{(2)}\rangle +
     \langle\mathbf{S}^{(2)} {\mathbf{\Delta}},\mathbf{\Delta}\mathbf{S}^{(1)}\rangle \right)\\
     &+
   \langle\mathbf{\Delta},\mathbf{S}^{(1)} - \mathbf{S}^{(2)}\rangle.
\end{split}    
\end{equation*}is a D-trace loss function, with $\langle A, B\rangle = \mbox{tr}\left(AB^\top\right)$, $||A||_{1} = \sum_{i, j = 1}^p|a_{ij}|$, and $\lambda > 0$ a tuning parameter. 

Similarly to the fused graphical lasso, the authors suggest to use alternating direction method of multipliers to solve the problem and obtain $\hat{\mathbf{\Delta}}$. 

We used the implementation of the method from the \texttt{DiffGraph} R package available on GitHub (not developed by the authors of the method, see software article \citep{DiffGraph_package}).

\bigskip

\section{Simulation study}\label{sec: results}
\subsection{Study set-up}

In this section, we introduce our simulation study set-up. To assess the methods' performance, we decided to vary the structure of the differential network, the sparsity of the graphs, and the  sample size. We elaborate on parameter configurations below.

\subsubsection{Graph generation}

In this study, we would like to investigate to what extent the structure of the differential network influences the methods' performance. It is well-known that presence of hubs (vertices with much higher degrees than the rest of the nodes in the graph) poses a challenge in estimation of graphical models \citep{MingTan2014LearningHubs} but, to the best of our knowledge, there are no studies on its effect in differential network estimation. However, in biological networks, hubs are likely to be present and efficient estimation of graphs with hubs is of utmost importance.

\begin{figure*}
\centering
  \includegraphics[width = 0.85\textwidth]{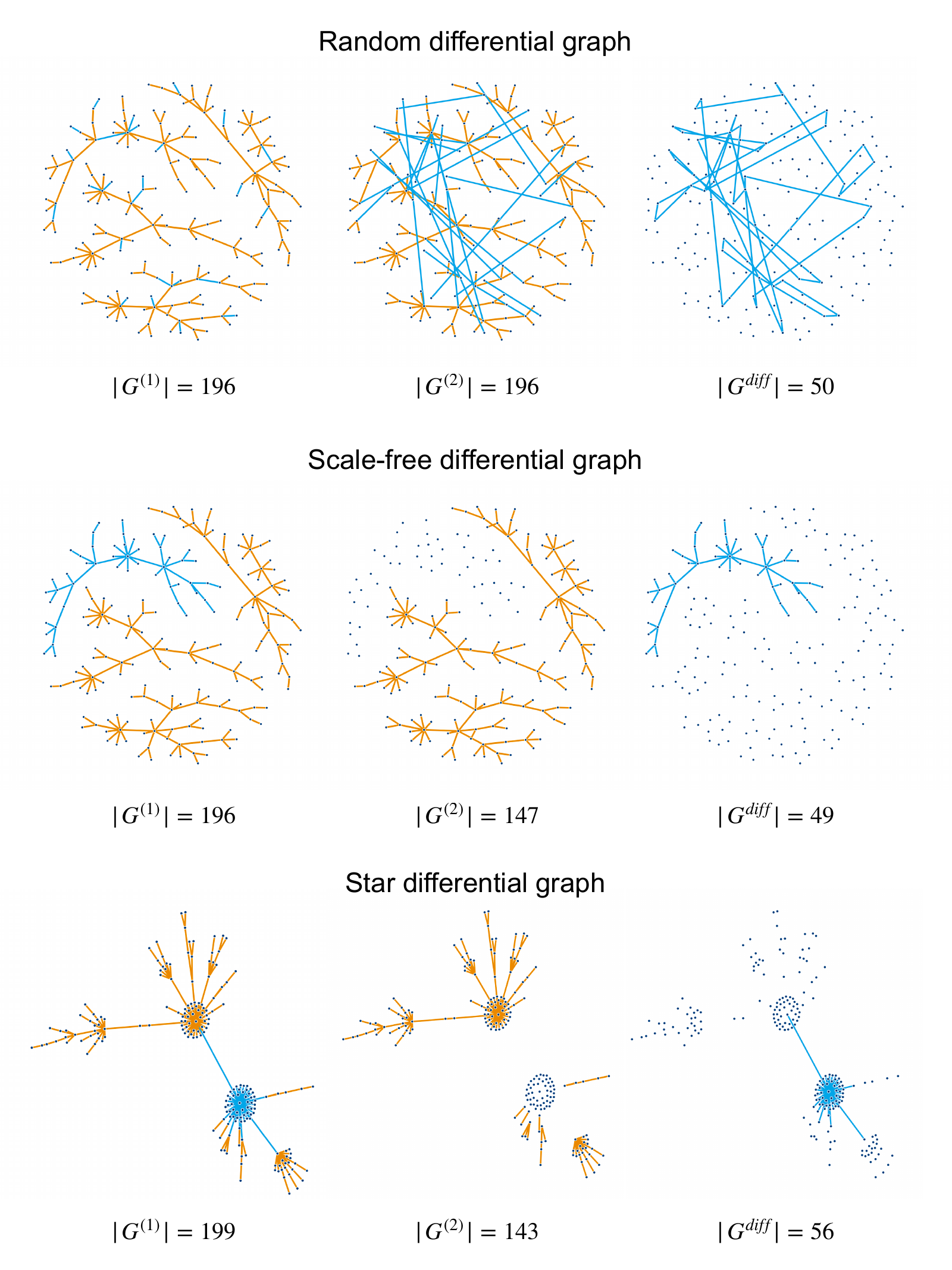}
  \caption{Graphs used in the simulation study for $|G^{(1)}| \approx 200$ and $|G^{diff}| \approx 50$ and 3 differential graph structures. Orange edges are common for $G^{(1)}$ and $G^{(2)}$; blue edges are differential, i.e. present only in $G^{(1)}$ or $G^{(2)}$.}
\label{fig_graphs_simulation}
\end{figure*}

We consider three different structures of $G^{diff}$(see  Figure \ref{fig_graphs_simulation}):

\begin{enumerate}
    \item\textbf{Random graph}: Start with a network $G^{(1)}$ that consists of several disconnected scale-free networks, i.e. its adjacency matrix is block-diagonal. Then, apply an iterative rewiring procedure. It randomly chooses pairs of edges $(a, b)$ and $(c, d)$ and switches them to obtain $(a, c)$ and $(b, d)$ edges. This produces a graph $G^{(2)}$ with the same size and degree distribution. The differential network $G^{diff}$ has no prominent structure. This set-up acts as a baseline where standard methods are assumed to work well.
    \item\textbf{Scale-free graph}:  Start with the same network $G^{(1)}$ consisting of several disconnected scale-free networks. Remove some blocks (the number depends on what sparsity we want to achieve) and obtain $G^{(2)}$. The differential network $G^{diff}$ consists of one or several disconnected scale-free networks. This case mimics the turn-off of a whole group of genes, with some genes being hubs. A similar set-up is used in the simulation studies in \citep{JGL} and \citep{Dtrace}.
    \item \textbf{Star/hub graph}: Start with a scale-free network $G^{(1)}$. Then, identify a hub (or two, depending on the desired sparsity  level) – a vertex with the biggest degree. Remove all edges, incident to that vertex, and obtain a graph $G^{(2)}$. The differential network $G^{diff}$ is therefore a star graph. This case mimics an inhibition of a gene. The simulation studies of \citep{TestingPartCorr} and \citep{DCA} have similar set-ups, although with a higher number of less prominent hubs.
\end{enumerate}

For all graph structures, there are several ways to vary the sparsity of $G^{(1)}$ and  $G^{diff}$:

\begin{enumerate}
    \item Fix $|G^{diff}|$ and vary the size of the \textbf{whole condition-specific} graph $|G^{(1)}|$;
    \item Fix $|G^{(1)}|$ and vary the size of the \textbf{differential graph}  $|G^{diff}|$;
    \item Fix the proportion $|G^{(1)}| /|G^{diff}|$
    and vary \textbf{both graphs'} sizes simultaneously.
\end{enumerate}

In this work, we simulate all these three possibilities to assess the influence of a graphs' sparsity on the methods' performance.

All graphs have $p = 200$ vertices. The number of edges (size) of $G^{(1)}$ is determined by the number of edges added at each step of the Barabasi-Albert algorithm for scale-free graph generation, one or two. The size of $G^{diff}$ and, consequently, $G^{(2)}$ is determined either by the rewiring procedure or the edge removal procedure. Since the graph generation steps are non-deterministic, they were re-run to achieve the best size consistency across different graph types and sparsity set-ups. The resulting graph sizes are given in Table \ref{table_graphs_sizes}. For simplicity, in the rest of the paper we will refer to the rounded-up graph sizes (50-200, 100-200, 50-400 and 100-400, respectively). We describe the graph generation procedure in detail in Appendix \ref{appendix_simulation}. See examples of generated graphs for the 50-200 edge setting in Figure \ref{fig_graphs_simulation} and all graphs in Figure (a) of Supplementary materials.

\begin{table}[h]
\centering
\caption{\label{table_graphs_sizes}Exact number of edges   $(G^{(1)}$, $G^{(2)},G^{diff})$ for different approximate sizes of $G^{(1)}$ and $G^{diff}$, and different graph types.}
\begin{tabular}{r|l|rr}
$|G^{(1)}|$ & Graph type & \multicolumn{2}{c}{$|G^{diff}|$}\\\hline
&& \multicolumn{1}{c}{$\approx  50$} & \multicolumn{1}{c}{$\approx 100$}\\
& Random & $(196,196,50)$ &$(196,196,98)$\\
$\approx 200$ & Scale-free & $(196,147,49)$ & $(196,98,98)$\\
& Star & $(199,143,56)$ & $(199,103,96)$\\
& Random & $(388,388,50)$ &$(388,388,100)$\\
$\approx 400$ & Scale-free & $(388,333,55)$ & $(388,291,97)$\\
& Star & $(397,342,55)$ & $(397,292,106)$\\
\end{tabular}
\end{table}

In the Discussion section, we will discuss properties of the pairs $(G^{(1)},G^{(2)})$ and the corresponding differential graphs in connection with varying performance.

\subsubsection{Data generation}

For all graphs, we use the same data generating procedure. To construct precision matrices $\mathbf{\Omega}^{(1)}$ and $\mathbf{\Omega}^{(2)}$, we start with the graphs' adjacency matrices. We keep zeroes  and  replace non-zero entries, corresponding to network edges, with values sampled randomly from the uniform distribution with support on $[-0.9, -0.6] \cup [0.6, 0.9]$. Entries corresponding to the common edges  in $\mathbf{\Omega}^{(1)}$ and $\mathbf{\Omega}^{(2)}$ are  the same. We ensure symmetry by construction, i.e. we put the same value for $(i,j)$ and $(j,i)$ elements. To ensure positive definiteness (unless already achieved by construction), we make the matrices diagonally dominant by setting diagonal elements to $\Omega_{ii}^{(k)} = \sum_{j \ne i} |\Omega_{ij}| + 0.1$. We then invert each precision matrix and obtain $\mathbf{\Sigma}^{(1)}$ and $\mathbf{\Sigma}^{(2)}$. Finally, we sample $n_1 = n_2 = n$ independent identically distributed observations to obtain $\mathbf{X}^{(1)}$ and $\mathbf{X}^{(2)}$. 

Two different sample sizes, $n_1 = n_2 = 100$ and $n_1 = n_2 = 400$, were chosen to reflect situations with $n < p$ and $n > p$. For biological applications, e.g. molecular networks, sample size is typically limited and higher sample sizes are not realistic for most studies. However, since we are estimating a graph, the number of parameters to estimate is not $p$ but $p(p - 1) / 2$, which is an additional challenge. Although it is known that the number of samples influences the performance, we aim to study the magnitude of its impact. 

 For each pair $(G^{(1)}, G^{(2)})$, we generate 50 datasets and report all performance results as an average over the 50 sets.

\subsubsection{Performance evaluation}

For each method we vary either the regularization parameter $\lambda$ (or $\lambda_1$ and $\lambda_2$) or the FDR level $\alpha$ to vary the size of  $\hat{E}^{diff}$, from almost complete to almost empty. For each value of the parameter, we compare the estimated differential graph with the true differential graph and calculate the number of true and false positive edges as 
\begin{equation*}
\begin{split}
TP &= | \{(i, j): (i, j) \in E^{diff} \cap \hat{E}^{diff}\}|, \\
FP &= |\{(i, j): (i, j) \in \left(E^{diff}\right)^c \cap  \hat{E}^{diff}\}|.
\end{split}
\end{equation*}
For each method and each parameter value, we use $TP$ and $FP$ to calculate empirical power (how large proportion of the true edges a method was able to identify) and empirical false discovery rate (proportion of false edges among all identified) as
\begin{equation*}
\begin{split}
\mbox{power} &= TP / |E^{diff}|,  \\
\mbox{FDR} &= FP / \max(1,FP + TP).
\end{split}
\end{equation*}

Considering these metrics across a range  of tuning or regularization parameters for each method, we obtain a curve – see Figure \ref{fig_perf_all}.

Values of the parameters used in the simulation study are reported in Appendix \ref{appendix_simulation_settings}. Simulation study code is available at \url{github.com/annaplaksienko/Diff_networks_review_simulation}.

\subsection{Results}

\begin{figure*}[h!] 
\includegraphics[width = 0.9\textwidth]{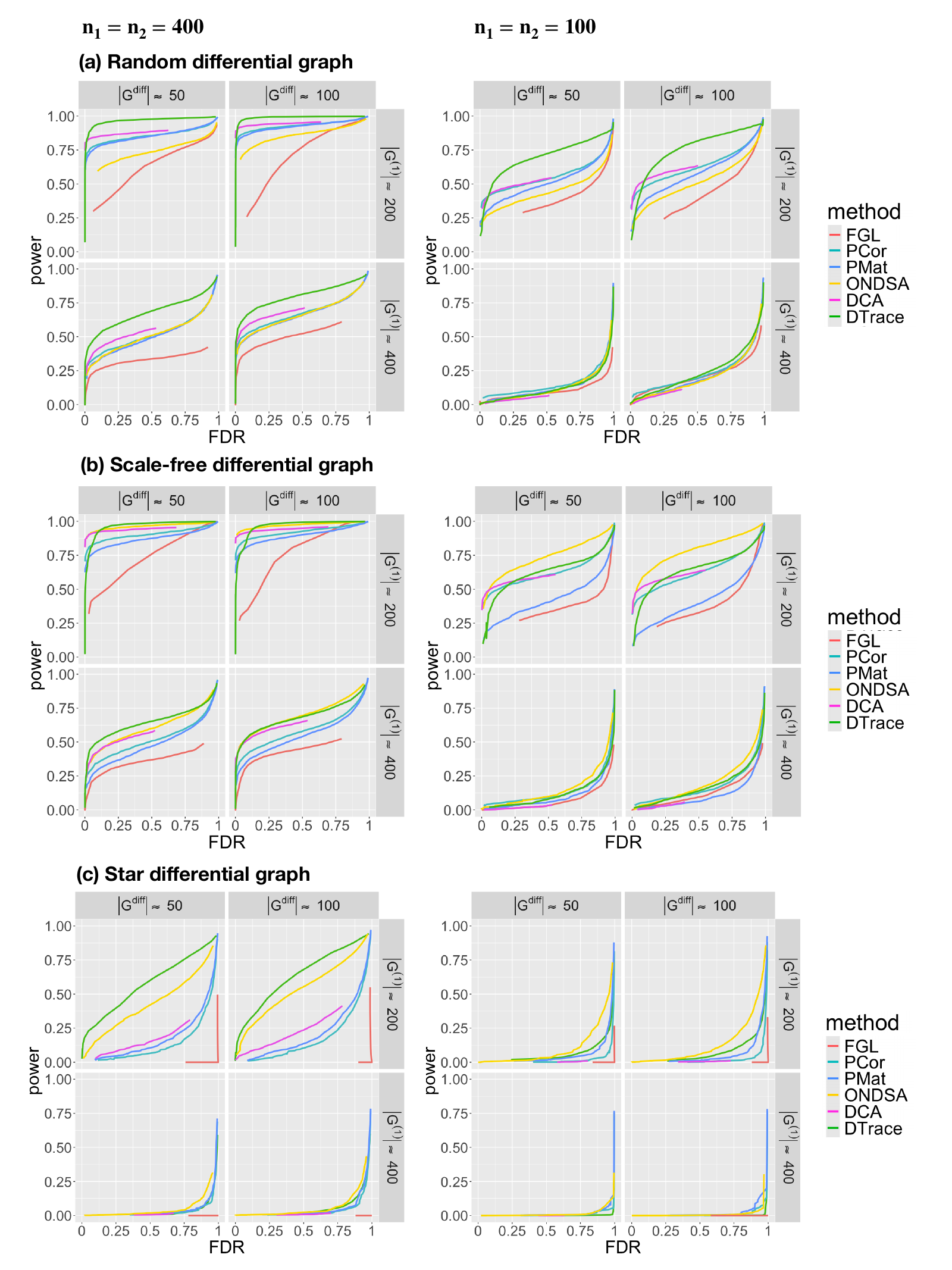}
  \caption{Estimated power vs FDR for the six considered methods in various sparsity and differential graph structure settings, with varying sample size. Results are averaged over 50 realizations of data.}
\label{fig_perf_all}
\end{figure*}

In Figure \ref{fig_perf_all}, we present the obtained results in terms of power versus false discovery rate for various methods (different colours), sparsity settings (sets of four facets in each panel of the figure), sample sizes (left and right set of facets in each panel) and differential graph structures (panels (a), (b) and (c)). We observe several challenges here.

For most settings, all methods exhibit the same trends depending on the graph structure and sparsity. Comparing various differential graph structures (see panels (a), (b) and (c) of Figure \ref{fig_perf_all}), we observe that overall, performance is similar for the random and the scale-free graphs. However, for the star differential graph the power is very low for almost all methods in all settings. In other words, the more "structure" there is in a differential graph, the worse the performance becomes. Moreover, we notice that in most cases, the size of $G^{diff}$ has little effect on performance (compare left and right facets in each panel) while an increase in size of $G^{(1)}$ from $\approx200$ to $\approx400$ edges influences performance dramatically (compare the first and the second row in each panel and notice the drop in performance). 

Regardless of the differential graph structure or size, we observe a significant deterioration of performance with a decrease from 400 to 100 samples (compare left and right set of facets in each panel of Figure \ref{fig_perf_all}), as expected. Although sample size thresholds depending on problem dimension $p$ and the number of non-zero elements (edges) exist for many sparse application problems (see \citep{Wainwright2009SharpLasso} for lasso, for example), to our knowledge, there are no results specific to differential graph estimation yet.

The best overall performance is demonstrated by DTrace (green line in the plots). It relies on direct estimation and focuses on the most distinct minimization problem. In most settings the difference in performance may not be large (although still noticeable), but in the 50-200 and 100-200 star differential graph setting it demonstrates a clear advantage compared to most other methods.

Testing methods PMat (blue line) and PCor (turquoise line) demonstrate very similar performance under most settings, with an exception of the scale-free differential graph, 50-200 and 100-200 edges with 100 samples, where PCor demonstrates noticeably superior performance. DCA (magenta line) usually is either on par with or slightly outperforms both. ONDSA (yellow line) is inconsistent, showing in some cases inferior performance to all above mentioned testing methods, and in some cases noticeably outperforming all of them – see scale-free differential graph, 50-200 and 100-200 edges with 100 samples and especially star differential graph, 50-200 and 100-200 edges with 400 samples.

For very low FDR, all testing methods often demonstrate better power than the above-mentioned DTrace method.

The joint estimation method FGL demonstrates the worst performance. In particular, for the star differential graph, this method has no power at all.

\section{Real data application}

Before moving on to the Discussion, we would like to also compare the presented methods on real data. This example, on the one hand, highlights the variability of results produced by different methods and hence emphasizes the difficulty of the problem, and on the other hand, provides evidence that the presented methods are capable of detecting known biological phenomena and should, therefore, be further developed and perfected.

\begin{figure*}[h!]
\centering
  \includegraphics[width = 0.95\textwidth]{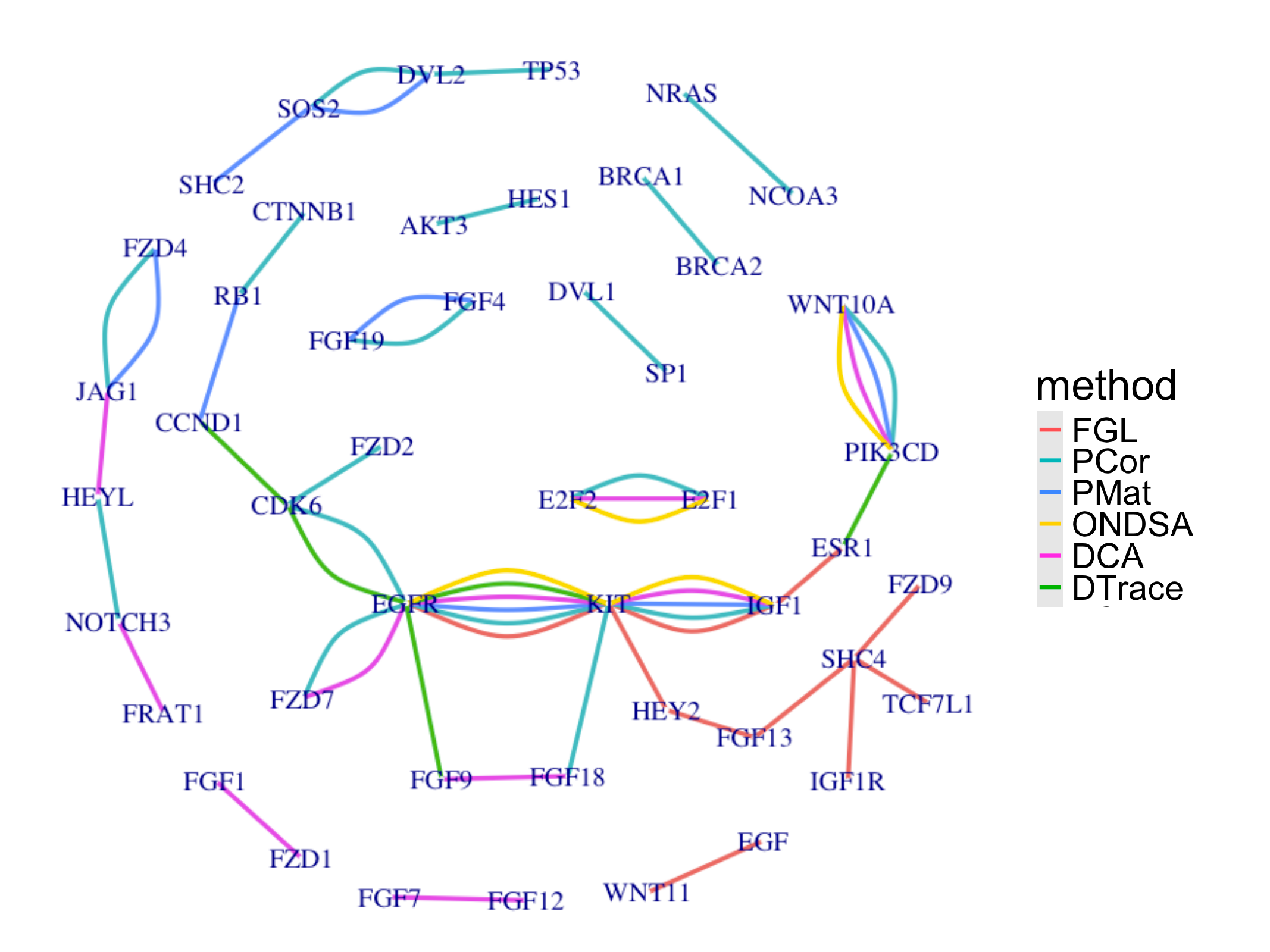}
  \caption{The figure demonstrates the variability of results produced with different methods. The multiplicity of an edge denotes how many methods have identified it and the color corresponds to a method. }
\label{fig_real_data}
\end{figure*}

We consider a dataset from \citep{BreastCancerRealData}, where the authors collected 526 samples of patients with breast cancer of four molecular subtypes: luminal A, luminal B, basal-like, and HER2-enriched cancers. The original microarray gene expression dataset has 17 327 genes. For this analysis, we use a pre-processed sub-dataset available in the \texttt{DiffNetFDR} R package \citep{DiffNetFDR_package}. It consists of data from two of the subtypes, the luminal A subtype with $n_1 = 231$ samples and the basal-like subtype with $n_2 = 95$ samples, and $p = 139$ genes from the hsa05224 KEGG \citep{KEGG} breast cancer pathway. 

To identify gene connections that differ between luminal A and basal-like subtypes of breast cancer, we used all the six differential network estimation methods that we discussed above. For the testing methods, we used  $\alpha = 0.05$, while for FGL and DTrace, the tuning parameters were estimated with the Bayesian Information Criterion (BIC). All values are provided in Appendix \ref{appendix-real-data}. The different methods produced networks of the following sizes (in decreasing order): PMat – 18 edges, DCA and FGL – 10 edges, PCor – 8 edges, DTrace – 5 edges, and ONDSA – 4 edges.

To compare the results, we constructed a union of all six estimated networks. Out of the original 139 vertices (genes), 92 are estimated to be isolated by all the methods. The size of the union of all six networks is 36 edges. Interestingly, only one edge was identified by all six methods (EGFR-KIT), by five out of six (KIT-IGF1) and by four out of six (Wnt10a-PIK3CD), respectively. Most of the edges were estimated by one method only (see Table \ref{table_real_data} and Figure \ref{fig_real_data}). 

\begin{table}[h] 
\caption{A frequency table showing the number of methods, $m$, whose estimated differential networks have $l$ common edges, i.e. one edge was present in all six estimated differential networks, one edge was present in five networks, one edge was estimated by four methods \ldots}
\label{table_real_data}
\centering
\begin{tabular}{ l|ccccc c| } 
 Number of methods $m$ & 6 & 5 & 4 & 3 & 2 & 1 \\ 
 Number of edges $l$ & 1 & 1 & 1 & 1 & 5 & 27 \\ 
\end{tabular}
\end{table}

The only edge estimated by all six methods is the EGFR-KIT edge. The authors of \citep{EGFR_KIT_1} and \citep{EGFR_KIT_2} report that expression of both of these genes is associated with the basal-like breast cancer subtype which suggests that this edge is a true differential edge, i.e. present in the basal-like subtype but not in luminal A. If we consider each node individually, \citep{KIT_basal_1} identifies c-KIT as a potential biomarker and possible molecular target for therapy for patients with basal-like breast cancer. For EGFR, also associated to the basal-like subtype, there has been arguments regarding if it is a potential therapeutic target \citep{EGFR_basal_2}
or "just" a biomarker \citep{EGFR_basal_1}.

We were not able to find specific literature support for the KIT-IGF1 edge, the edge estimated by five out of six methods. However, we already mentioned above that c-KIT has been associated with basal-like breast cancer and similar evidence can be found for IGF1. The authors of \citep{IGF1_basal_2} mention that basal-like tumors frequently induce IGF1 transcriptional upregulation and studies \citep{IGF1_basal_1, IGF1_basal_3} identify IGF1 as a therapeutic target specifically in basal-like breast cancer. 

The edge Wnt10a-PIK3CD was estimated by four out of six methods. In \citep{WNT10A_basal} it was discovered that Wnt10a is strongly linked to the basal-like subtype, while the authors of \citep{PIK3CD_luminal} hypothesized that PIK3CD is associated with a favorable prognostic biomarker for overall survival in the luminal A subtype. 

We consider it promising that edges with known biological support were identified by all/most methods, even if some evidence is circumstantial. Regarding the "single" edges, i.e. those identified only by one of the methods, it is unclear whether those are false positives or if different methods capture different parts of the true graph.

\section{Discussion}

We have investigated the performance of six different methods for differential network estimation under varying graph structure, sparsity and sample size. We find direct estimation with lasso penalized D-trace loss to perform the best overall. However, given our focus on structure learning, one clear drawback of this method is the lack of formal error control. This is an area for future research. 

The worst performance was shown by the joint estimation method FGL. This is not so surprising, as the method is constructed for an entirely different purpose. The method has two tuning parameters. One of them is driving sparsity, while the other one is targeted towards the degree of similarity between the precision matrices, not differences. Moreover, the necessity to tune two parameters is in itself a drawback compared to the other methods. This problem may be solved as suggested in \citep{StabJGL} with a combination of stability selection and an extended Bayesian information criterion.

Regarding error control, PMat, PCor and ONDSA controlled FDR at the nominal level $\alpha$, with the exception of the star graph settings, where the methods were much less stable and occasionally empirical FDR exceeded $\alpha$ (see Figure (b) in the Supplementary materials). For ONDSA, FDR was also not controlled in the 50-200 random graph setting with 400 samples. The reason for this is not clear. Regarding the DCA, it was overall more conservative. However, in some settings empirical FDR significantly exceeded the target level for small values of $\alpha$. The inflation of FDR occured in general for the smaller sample size and was even more pronounced for the star differential graph. This finding is not surprising since error control for this method strongly relies on the coverage property, that is, on the event that the estimated neighborhood of each node covers its true neighborhood. This event has smaller probability both when $n$ is small and when the size of the neighborhood is large, as is in a star graph.

We found that overall performance was primarily related to graph structure. We would like to investigate this a bit further.

We expect that performance of methods for learning differential networks depends on various properties of underlying graphs and probability distributions. To gain further insight, various measures of these properties can be employed. We considered two measures: one that corresponds to a global distance between distributions $\mathcal{N}(0, \mathbf{\Sigma}^{(1)})$ and $\mathcal{N}(0, \mathbf{\Sigma}^{(2)})$, and one that summarizes the local structure of a differential graph.

We evaluated the global distance between distributions using symmetrized Kullback-Leibler divergence $KL_{12} + KL_{21}$, where for two centered Gaussian distributions  $\mathcal{N}(0, \mathbf{\Sigma}^{(i)})$, $i =1, 2$,  the Kullback-Leibler divergence $KL_{ij}$, $i, j = 1, 2$,  is defined as
\begin{equation*}
KL_{ij}  = \dfrac{1}{2}\left\{\mbox{tr}\left(\left[\mathbf{\Sigma}^{(j)}\right]^{-1} \mathbf{\Sigma}^{(i)}\right) - \log \dfrac{\det \mathbf{\Sigma}^{(i)}}{\det \mathbf{\Sigma}^{(j)}} - p\right\}.
\end{equation*}

To quantify local differences, we
used the highest vertex degree of a differential graph divided by the total number of vertices $p$. Note that normalization by $p$ is not necessary in the presented simulation as all graphs have the same size, $p = 200$, but we use it for a possible comparison with other studies.

\begin{figure*}[h!]
  \includegraphics[width = 0.95\textwidth]{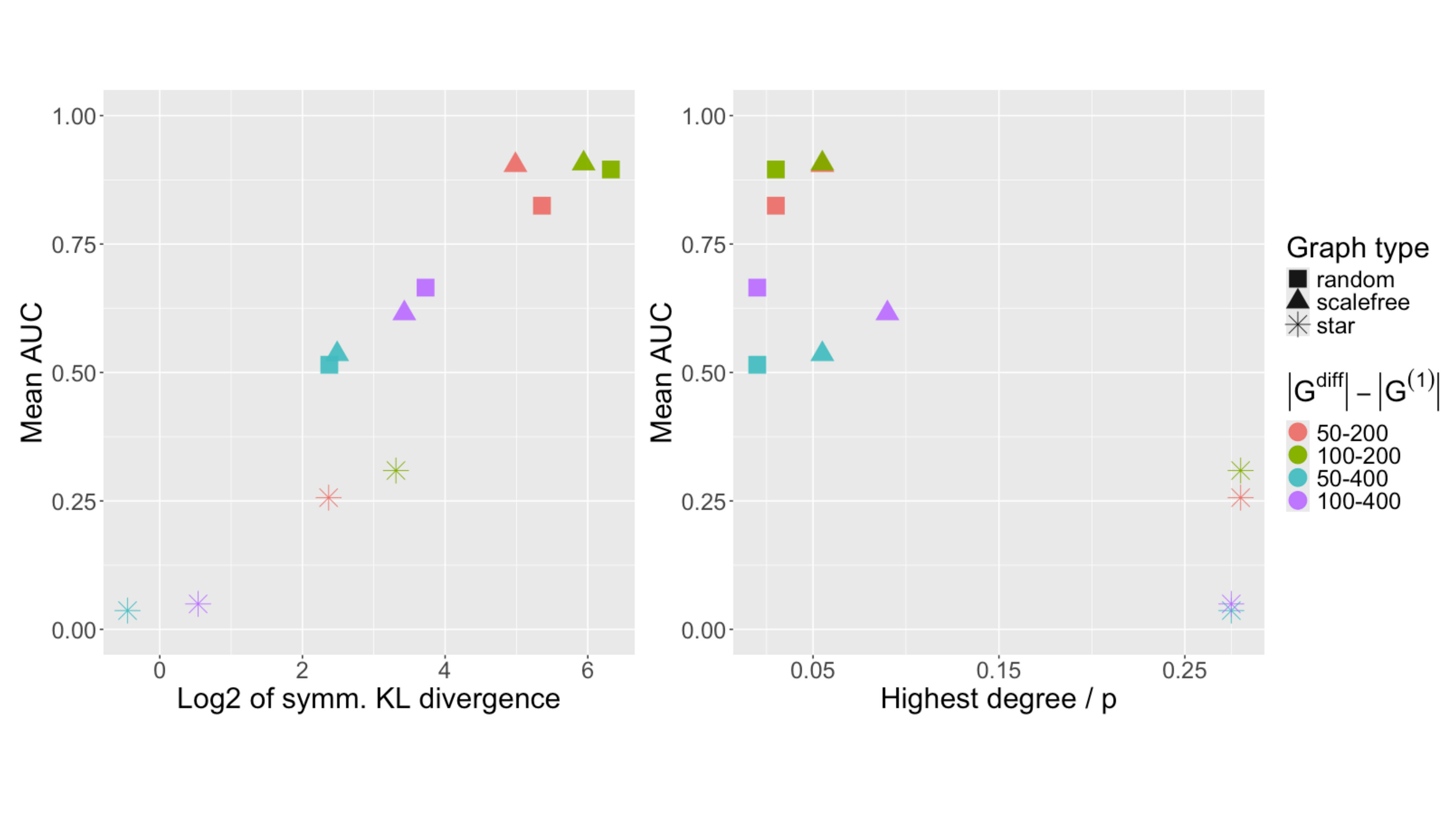}
  \caption{Averaged mean area under the power vs. FDR curve as a function of the symmetrized Kullback-Leibler divergence(left panel) and the highest vertex degree of a differential graph (right panel).}
\label{fig_distances}
\end{figure*}

And finally, to summarize the performance curves with a scalar, we computed the area under the curve (for power vs FDR) for  $n = 400$ samples and averaged it over all six methods. We related this performance indicator to the symmetrized Kullback-Leibler divergence and to the highest vertex degree. The resulting plots are shown in Figure \ref{fig_distances}.

The left panel of Figure \ref{fig_distances} demonstrates that methods on average perform better for sufficiently distant distributions (e.g. the cases of random and scale-free graphs in 50-200 and 100-200 edges settings). In other words, the larger the global differences are, the easier it is to estimate them correctly. 

We notice that for the star differential graph, in some settings, the distance between distributions is as high as for random and scale-free graphs, but the performance of the methods is still considerably worse, measured by mean AUC. This may be explained by the local structures, as illustrated by the normalized highest degree, that is high for the star graph in all settings. It is well known that hub structures are harder to estimate, as most methods assume a uniform distribution of edges, without any specific structure (see discussion in \citep{MingTan2014LearningHubs}, for example). Therefore, more information is required for a successful estimation in these situations. Indeed, for one of the methods (PMat, since it is one of the fastest) we have evaluated the performance for higher sample sizes, $n_1 = n_2 = 1000, 2000$, and discovered a considerable improvement (see Figure (c) in the Supplementary materials). We hypothesize this observation applies  to other methods as well. Although increased sample size may be unachievable in many situations, aiming for as many samples as possible is advisable. 

In the absence of larger samples, another option for improving performance in situations with hubs is to incorporate prior hub knowledge in the analysis. Such knowledge is often available, e.g. in the context of certain types of molecular networks. For instance, the joint estimation method \textit{jewel} \citep{jewel} is using user-provided class-specific degrees to construct edge weights which are then incorporated into the penalty terms for the method.

Methods for learning individual networks with hub structure do also exist, see e.g. the hub graphical lasso \citep{MingTan2014LearningHubs} or the hub weighted graphical lasso \citep{Hub_weighted_glasso}. Methods such as the perturbed node graphical lasso \citep{PNJGL} estimate two individual networks jointly taking into account potential hub differences. A possible direction of future work could be to extend such methods to the differential network situation, e.g. by using their estimands as an input for the testing-based methods described in the current paper.

In the current work we focused on networks with hubs since they often find application in modeling of biological networks. However, we expect that the observed challenges for the existing statistical methods carry over to more general community structures. 

Finally, related to the sample size, it would be desirable to obtain theoretical results on the necessary sample size $n$ in relation to the differential graph order $p$ and its maximum degree, as done, for example, for Ising models \citep{Ising_Models_sample_size}.

\section*{Declarations}

\subsection*{Author contributions}

AP, MT and VD conceived the research idea. AP conducted the simulations and the real data analysis, and drafted the manuscript with input from MT and VD. All authors gave final approval.

\subsection*{Funding}

This work is supported in part by the Norwegian Cancer Society, project number 216137.

\subsection*{Data availability}

The raw breast cancer data used in our motivating example is available at \url{https://gdc.cancer.gov/about-data/publications/brca_2012} as \textit{BRCA.exp.547.med.txt} (microarray gene expression) and \\ \textit{BRCA.547.PAM50.SigClust.Subtypes.txt} (cancer subtypes). The pre-processed dataset we used is a part of \texttt{DiffNetFDR} package \cite{DiffNetFDR_package} as a \texttt{TCGA.BRCA} object.

\subsection*{Code availability}

Simulation study code is freely available at \url{github.com/annaplaksienko/Diff_networks_review_simulation}.

\begin{appendices}

\section{Real data settings}\label{appendix-real-data}

Here, we provide the parameters used to estimate a networks with optimal parameters in the real data example. 

\begin{itemize}
    \item For fused graphical lasso (FGL) \citep{JGL}, implemented in the package \texttt{JGL}, we first fix $\lambda_2 = 0.01$ and chose $\lambda_1$ with BIC, obtaining $\lambda_1 = 0.55$. Then, using that value, we perform search over the grid for $\lambda_2$. As we fail to find BIC minimum inside the interval, we propose to chose a parameter corresponding to an "elbow", i.e. slight change in trend of the BIC curve. That is $\lambda_2 = 0.286$. We select a differential edge if $|\hat{\Omega}_{ij}^{(1)} - \hat{\Omega}_{ij}^{(2)}| > 10^{-3}$ (as the authors do in the code accompanying their manuscript);
    \item For the PCor method \citep{TestingPartCorr}, implemented in the package \texttt{DiffNetFDR} \citep{DiffNetFDR_package}, we set $\alpha = 0.05$;
    \item For the PMat method \citep{TestingPrecMatr}, implemented in the package \texttt{DiffNetFDR} \citep{DiffNetFDR_package}, we set $\alpha = 0.05$;
    \item For omics networks differential and similarity analysis (ONDSA) \citep{ONDSA} implemented in the package \texttt{ONDSA}, we set $\alpha = 0.05$;
    \item For the differential connectivity analysis method \citep{DCA}, implemented in the package \texttt{DCA}, we set $\alpha = 0.05$;
    \item For the DTrace method \citep{Dtrace} implemented in the package \texttt{DiffGraph} \citep{DiffGraph_package}, we chose optimal parameter with BIC and obtained $\lambda = 0.62$.
\end{itemize}

\section{Simulation study data generation}\label{appendix_simulation}

As described in the article, we have three differential graph structures: random, scale-free, and star. Let us first focus on the first two, random and scale-free, when $|G^{(1)}| \approx 200$. We start with the same graph $G^{(1)}$. We use \texttt{barabasi.game} from the \texttt{igraph} package to generate four disconnected scale-free networks, each with the power of preferential attachment (parameter \texttt{power}) equal to 1 and the number of edges added at each step of the generative algorithm equal to 1 (parameter \texttt{m}). The resulting graph $G^{(1)}$ has 196 edges. Then, to obtain a scale-free differential network, we remove either one or two scale-free networks, resulting in graphs $G^{diff}$ of sizes 50 or 98. To obtain a random differential network, we use a rewiring procedure that is performed with the \texttt{rewire(..., keeping\_degseq())} function from the \texttt{igraph} package with a parameter \texttt{niter} set to 15 or 30 and re-run several times to obtain $G^{diff}$ of sizes 49 and 98. 

The procedure is the same when $G^{(1)} \approx 400$, but now each of the four disconnected scale-free networks has $m = 2$ edges added at each step of the generative procedure, resulting in a subgraph of $\approx 100$ edges. To obtain a random differential network in the 50-400 and 100-400 settings, we use the same \texttt{rewire(..., keeping\_degseq())} function with a parameter \texttt{niter} set to 30 or 60 and re-run several times until we achieve the desired differential network size. To obtain a scale-free differential network in the 100-400 setting, we remove one of the disconnected subnetworks of $G^{(1)}$. To obtain $G^{diff}$ in the 50-400 setting, we construct a subgraph of that disconnected subnetwork in the following way: going from the vertex with the highest degree to the vertex with the lowest degree, we remove half of the incident edges of that vertex, controlling at each step that the number of connected components of $G^{diff}$ is equal to 1 (except for isolated vertices). This is done to preserve the scale-free structure as much as possible. 

Next, we will describe the generation of the star differential graph. To obtain the 50-200 and 100-200 settings, we first re-run the \texttt{barabasi.game} function with \texttt{m = 1} and \texttt{power = 1.7} until we obtain a graph that has two "hubs" (vertices with the highest degrees) of a similar size, 56 and 48. Then we remove all edges incident to one of them to obtain a 50-200 setting (the differential graph is a star graph) and all edges incident to both of them to achieve a 100-200 setting (the differential graph consists of two star graphs connected with one edge). To obtain the 50-400 and 100-400 settings, we re-run the \texttt{barabasi.game} function with \texttt{m = 2} and \texttt{power = 1.5} until we obtain a graph that has two "hubs" of similar size, 55 and 51. We then repeat the procedure described above.

Plots of all generated graphs can be found in Figure (a) of Supplementary materials. Code used to generate all graphs is freely available at \url{github.com/annaplaksienko/Diff_networks_review_simulation}

\section{Simulation settings}\label{appendix_simulation_settings}

Here we provide the parameters used to construct the power vs FDR curve in our simulation studies. We used the same set of parameters to produce the curves in all graph settings. We used default convergence parameters for all methods.

\begin{itemize}
    \item For fused graphical lasso (FGL) \citep{JGL}, implemented in the package \texttt{JGL}, we fix the "sparsity" tuning parameter $\lambda_1 = 0.1$ as we observed it provides good performance and we vary the "similarity" tuning parameter $\lambda_2$ from 0.01 to 0.5. We define an edge $(i, j)$ as differential if $|\hat{\Omega}_{ij}^{(1)} - \hat{\Omega}_{ij}^{(2)}| > 10^{-3}$, following authors code, accompanying their paper;
    \item For the PCor method \citep{TestingPartCorr}, implemented in the package \texttt{DiffNetFDR} \citep{DiffNetFDR_package}, we vary FDR level $\alpha$ from 0.001 to 0.995;
    \item For the PMat method \citep{TestingPrecMatr}, implemented in the package \texttt{DiffNetFDR} \citep{DiffNetFDR_package}, we vary FDR level $\alpha$ from 0.001 to 0.995;
    \item For the differential connectivity analysis method \citep{DCA} with default GraceI test, implemented in the package \texttt{DCA}, we vary FDR level $\alpha$ from 0.0001 to 0.95;
    \item For the DTrace method \citep{Dtrace},implemented in the package \texttt{DiffGraph} \citep{DiffGraph_package}, we vary the tuning parameter $\lambda$ from 0.1 to 0.7.
\end{itemize}

\end{appendices}

\bibliography{sn-bibliography}

\includepdf[pages = {1-3}, scale=0.9, angle=-270,  offset=0 0]{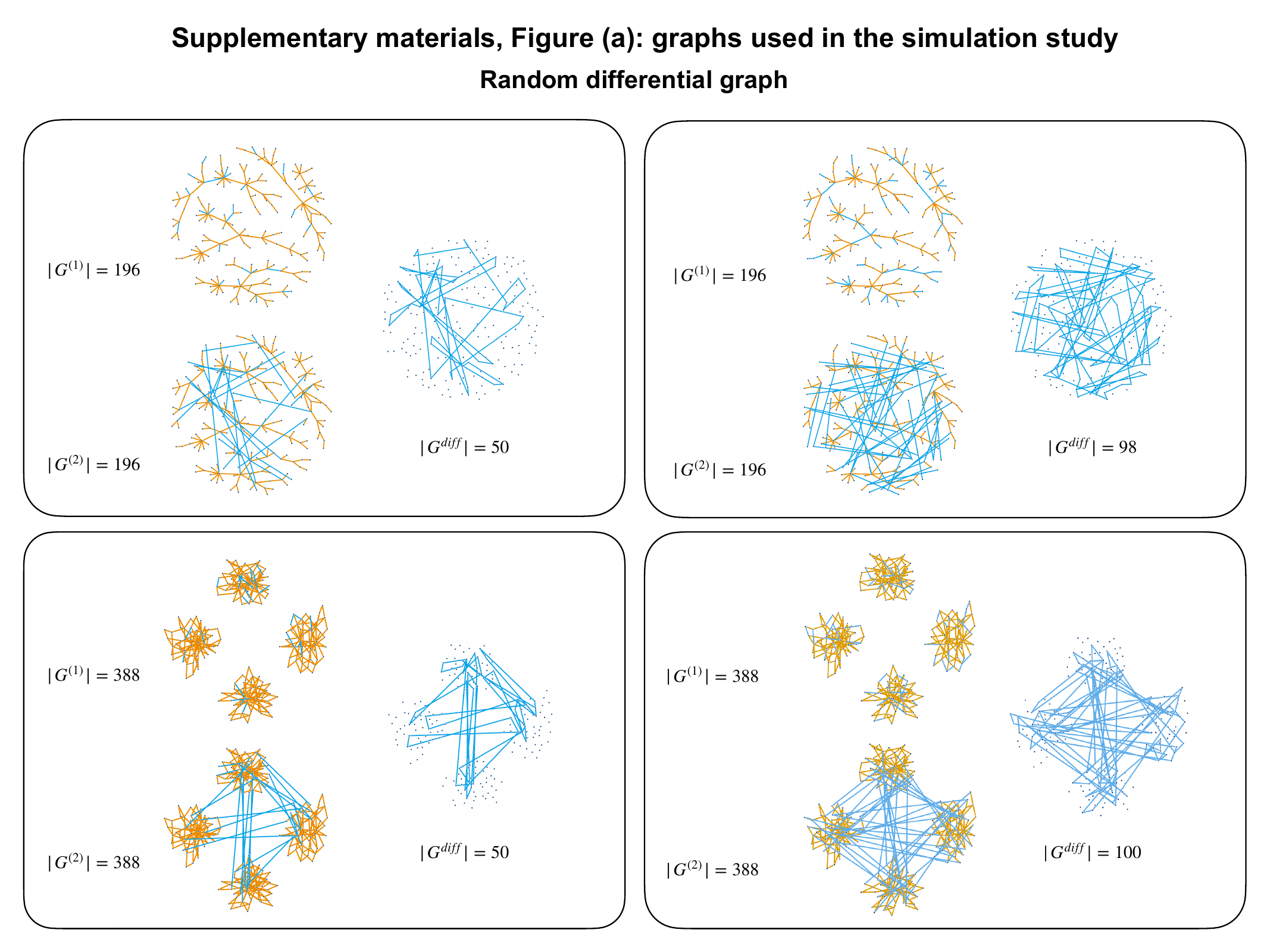}

\includepdf[pages = {4}, scale=0.9, offset=0 0]{Supplementary_abc.pdf}

\includepdf[pages = {5}, scale=0.9, angle=-270,  offset=0 0]{Supplementary_abc.pdf}

\end{document}